\documentclass[sigconf, authorversion]{acmart}

\usepackage{balance}
\usepackage{caption}
\usepackage{subcaption}
\usepackage{multirow}

\usepackage[version-1-compatibility]{siunitx}

\AtBeginDocument{%
  \providecommand\BibTeX{{%
    \normalfont B\kern-0.5em{\scshape i\kern-0.25em b}\kern-0.8em\TeX}}}

\copyrightyear{2022}
\acmYear{2022}
\setcopyright{rightsretained}
\acmConference[AutomotiveUI '22 Adjunct]{14th International Conference
on Automotive User Interfaces and Interactive Vehicular
Applications}{September 17--20, 2022}{Seoul, Republic of Korea}
\acmBooktitle{14th International Conference on Automotive User
Interfaces and Interactive Vehicular Applications (AutomotiveUI '22
Adjunct), September 17--20, 2022, Seoul, Republic of
Korea}
\acmDOI{10.1145/3544999.3552533}
\acmISBN{978-1-4503-9428-4/22/09}

\begin{document}


\title{In-Vehicle Interface Adaptation to Environment-Induced Cognitive Workload}


\author{Elena Meiser}
\affiliation{%
  \institution{German Research Center for Artificial Intelligence (DFKI)}
  \city{Saarbr{\"u}cken}
  \country{Germany}
}
\email{elena.meiser@dfki.de}

\author{Alexandra Alles}
\affiliation{%
  \institution{German Research Center for Artificial Intelligence (DFKI)}
  \city{Saarbr{\"u}cken}
  \country{Germany}
}
\email{alexandra_katrin.alles@dfki.de}

\author{Samuel Selter}
\affiliation{%
  \institution{German Research Center for Artificial Intelligence (DFKI)}
  \city{Saarbr{\"u}cken}
  \country{Germany}
}
\email{samuel.selter@dfki.de}

\author{Marco Molz}
\affiliation{%
  \institution{German Research Center for Artificial Intelligence (DFKI)}
  \city{Saarbr{\"u}cken}
  \country{Germany}
}
\email{marco.molz@dfki.de}

\author{Amr Gomaa}
\affiliation{%
  \institution{German Research Center for Artificial Intelligence (DFKI)}
    \city{Saarbr{\"u}cken}
  \country{Germany}
}
\affiliation{%
  \institution{Saarland Informatics Campus}
  \city{Saarbr{\"u}cken}
  \country{Germany}
}

\email{amr.gomaa@dfki.de}

\author{Guillermo Reyes}
\affiliation{%
  \institution{German Research Center for Artificial Intelligence (DFKI)}
    \city{Saarbr{\"u}cken}
  \country{Germany}
}
\affiliation{%
  \institution{Saarland Informatics Campus}
  \city{Saarbr{\"u}cken}
  \country{Germany}
}

\email{guillermo.reyes@dfki.de}

\renewcommand{\shortauthors}{Meiser et al.}

\begin{abstract}
Many car accidents are caused by human distractions, including cognitive distractions. In-vehicle human-machine interfaces (HMIs) have evolved throughout the years, providing more and more functions. Interaction with the HMIs can, however, also lead to further distractions and, as a consequence, accidents. To tackle this problem, we propose using adaptive HMIs that change according to the mental workload of the driver. In this work, we present the current status as well as preliminary results of a user study using naturalistic secondary tasks while driving (i.e., the primary task) that attempt to understand the effects of one such interface.
\end{abstract}




\maketitle

\section{Introduction}
Driving is one of the most complex everyday tasks, involving several visual and auditory mental subtasks~\cite{FASTENMEIER2007drivingtaskanalysis}. Driving environment complexity, in terms of visual complexity and vehicle control difficulty, affect mental workload (MWL)~\cite{FASTENMEIER2007drivingtaskanalysis, Verwey2000On-lineMeasures}. MWL is an interaction of task demands, other environmental factors, and human characteristics~\cite{Marquart2015ReviewWorkload, tao2019systematic} and refers to the fact, that drivers need to expend mental effort to maintain a safe driving behavior~\cite{boer2001behavioral}. High MWL, in particular overload, is associated with more driving mistakes and traffic accidents~\cite{paxion2013does}. Besides decrements in performance, MWL can also be assessed by physiological measurements like heart rate (HR) and heart rate variability (HRV)~\cite{cain2007review,gomaa2022s}. Those can be easily acquired while driving with non-invasive tools and can be used to classify the current MWL level~\cite{charles2019measuring,gomaa2022s}. 
In the last few decades, in-vehicle human-machine interfaces (HMI) have become increasingly advanced. They generally have a great impact on the relationship between the driving task and MWL~\cite{lansdown2004distraction}. On one hand, they can add comfort to the driving task or even aid in its execution, e.g. giving directions or warnings~\cite{vaezipour2015reviewing}. On the other hand, they can also be a source of distraction or additional task load, e.g., giving too much information~\cite{yang2020distraction}. In the first case, HMIs can help to reduce the MWL of the driver while in the latter case they add MWL, which could lead to driving mistakes or even accidents. Adapting the HMI to the current MWL of the driver or the difficulty of the driving environment could help reduce the MWL and prevent a mental overload~\cite{piechulla2003reducing}. This can be implemented by reducing the information available on the display with rising MWL~\cite{patzold2021adaptive}. 
Since this information reduction equals a change on the display, it could also be possible that drivers find that change itself distracting or irritating. In this study, we wanted to test if adapting the HMI to a difficult driving environment reduces MWL compared to a static display that keeps the available information constant. We measured MWL physiologically via HR and HRV as well as behaviorally via driving performance. Additionally, we assessed secondary task performance by letting participants interact with the HMI while driving. Since user experience (UX) of in-vehicle HMIs is an important aspect, we collected UX data via a questionnaire. 

\section{Research Question and Hypotheses}

In this work, we investigate the feasibility of adaptive versus static HMIs in terms of MWL reduction (indicated by driving performance, secondary task performance, and heart rate measurements) and UX enhancement. Therefore, we formulated four hypotheses as follows.

\begin{itemize}

\item\textbf{Hypothesis H1}: The increase in MWL with environmental difficulty is lower for an adaptive system compared to a static system.
    
\item\textbf{Hypothesis H2}: In the task conditions, there is a higher increase in MWL with environmental difficulty than in the no-task condition.

\item\textbf{Hypothesis H3}: The increase in MWL with environmental difficulty is moderated by task difficulty. For more difficult tasks, the increase is larger. 

\item\textbf{Hypothesis H4}: The UX of an adaptive system is better compared to a static system.
\end{itemize}

\section{User Study}

\subsection{Demographics}
To test our hypotheses, we conducted a user study using a driving simulator. The study sample consisted of 35 participants divided into two groups: adaptive \textit{n}=16 and static \textit{n}=19 participants. The static and adaptive groups do not differ significantly in most of the demographic variables, as well as in the motivation and perceived difficulty of the tasks. However, the static group drove significantly more kilometers (\textit{M} = 10639, \textit{SD} = 4259) in the past year (\textit{t}(33) = 2.35, \textit{p} = .025, \textit{d} = -0.82) compared to the adaptive group one (\textit{M} = 2733, \textit{SD} = 12835). 

\subsection{Driving Environment Design}
For the driving task, we utilized OpenDS~\cite{math2013opends}, a medium-fidelity driving simulator. We designed two different environments inspired by Nowosielski et al.~\cite{nowosielski2018good}. The two environments differed in terms of visual complexity and vehicle control difficulty. The easy environment consisted mainly of straight segments or wide curves on a country road with few peripheral visual stimuli, such as trees and bushes. The complex environment consisted of many tight curves and turns in a city environment with a lot of peripheral differing stimuli, namely high buildings. The experiment lasted approximately 30 minutes, started in the countryside, and ended in the city. Participants were told to drive according to everyday traffic rules while keeping a speed limit of 50 km/h. 
Both of the environments were separated into four segments by stop signs and 60-second breaks. 

\subsection{Interface Design}
\begin{figure}[t]
	\centering
    \begin{subfigure}[t]{0.46\textwidth}
         \centering
         \includegraphics[width=6cm,height=4cm,keepaspectratio]{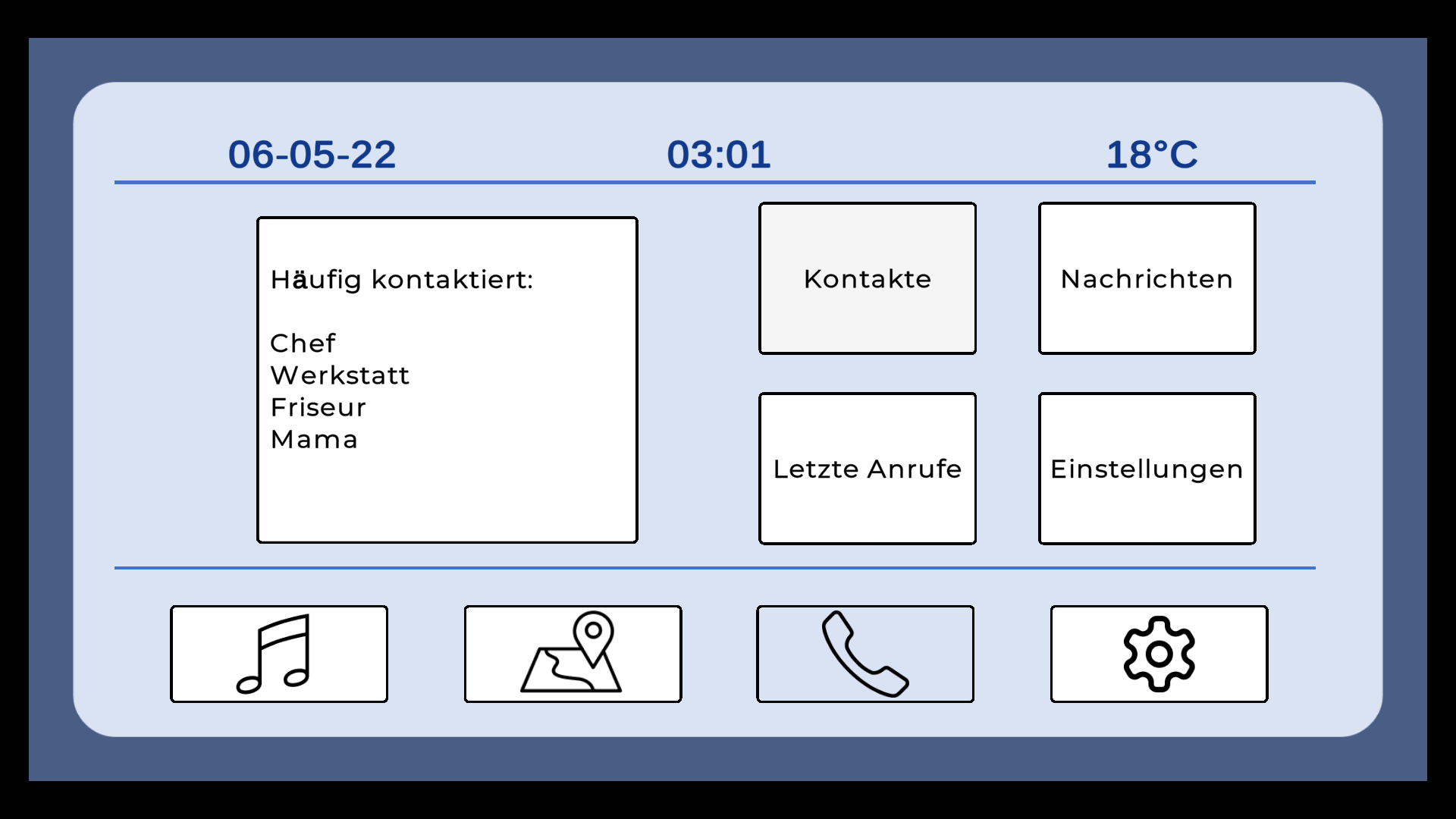}
         \caption{Complex Interface}
         \Description{Complex Interface}
         \label{fig:phone}
     \end{subfigure}
     \begin{subfigure}[t]{0.46\textwidth}
         \centering
         \includegraphics[width=6cm,height=4cm,keepaspectratio]{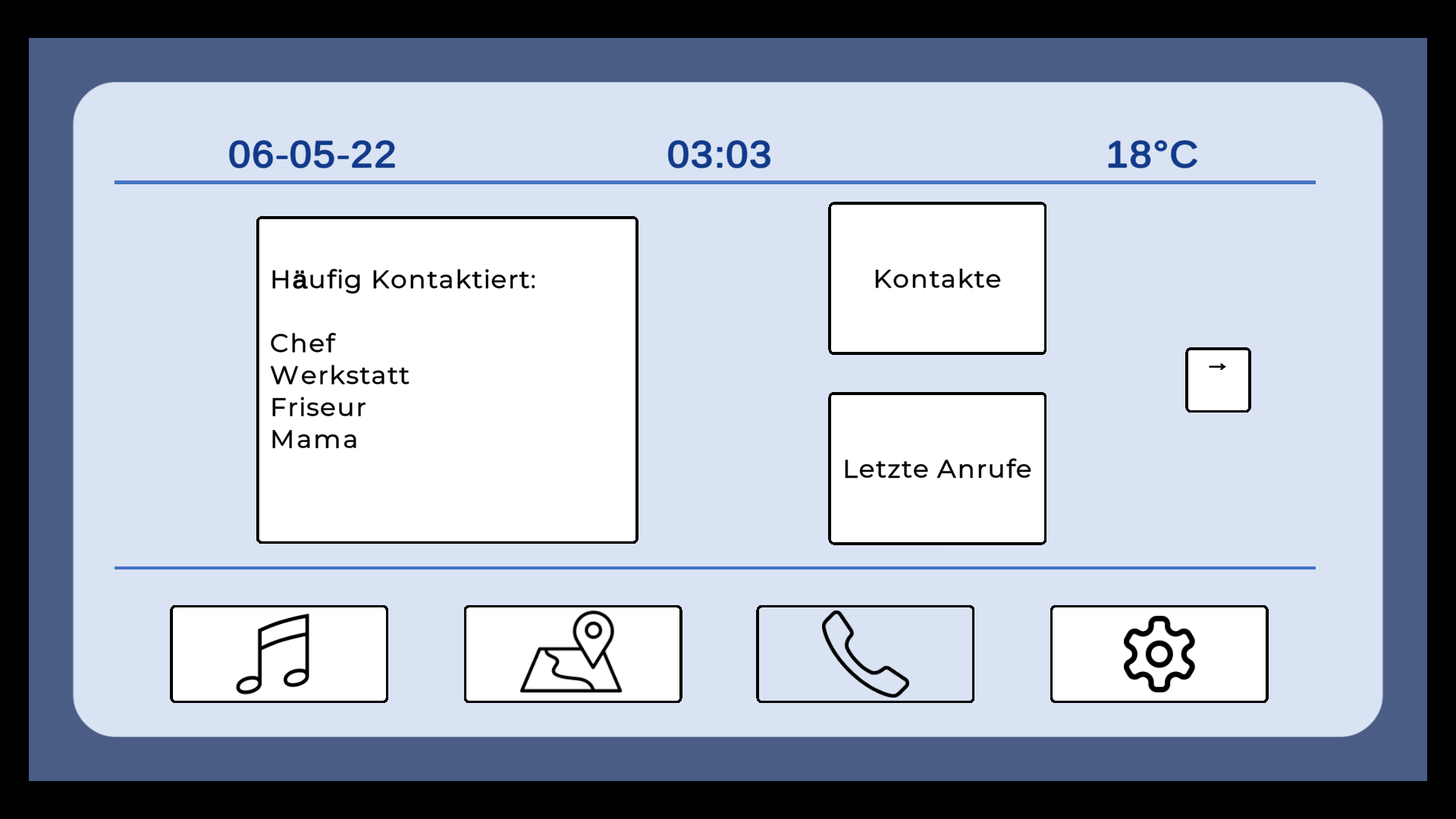}
         \caption{Simplified Interface}
         \Description{Simplified version of the interface. The interface shows a reduced amount of information in the form of buttons, yet the same functionality}
         \label{fig:phoneAdapt}
     \end{subfigure}
     \caption{(a) The complex interface (full information) and (b) the simplified interface (low information) of the phone menu. In the static condition, participants saw the complex interface in both environments while in the adaptive condition, they saw the complex interface in the countryside and the simplified interface in the city.}
     \Description{(a) The complex interface (full information) and (b) the simplified interface (low information) of the phone menu. In the static condition, participants saw the complex interface in both environments while in the adaptive condition, they saw the complex interface in the countryside and the simplified interface in the city.}
     \label{fig:intefrace_design}
\end{figure}

After literature research and a pre-study, we designed a a static interface and an adaptive one (see fig. \ref{fig:intefrace_design}). Three of the four menus were used (media, navigation, phone). For each of the menus, tasks of varying difficulty were designed. The easy tasks could be completed with one click on the interface and resembled menu switches or closing of pop-up information on the display. Medium tasks could be completed with four interactions (e.g., calling someone from contacts or choosing a navigation goal from the address book). The hard task could be completed with eight entries and included a search of a contact person, navigation goal, or a music playlist with four letters via a qwerty keyboard. Each of the environments consisted of four randomized segments: easy, medium, hard, or no-task conditions. To keep the number of interactions with the display as constant as possible, there were nine easy tasks triggered every ten seconds, six medium tasks triggered every fifteen seconds, and three hard tasks triggered every thirty seconds in their corresponding segment.

The adaptive interface differed only in terms of information presented on the display, while the functionality remained the same. Therefore, two pages with often used and less often used functions were designed. In terms of keeping the most used functions on the first page of each menu, the buttons needed to complete the tasks were always available on that page. This also ensured that the number of entries needed to complete the tasks was equal in both display designs.
Participants were assigned to the static or the adaptive condition, counterbalanced. To investigate if a display change facilitates or hinders the performance in the driving or the interface task, the display only changed for the adaptive condition entering the city, while it stayed the same for the static condition.

\subsection{Measurements}
All participants wore a commercial ECG-based sensor (Polar H10) attached to the chest, measuring HR and HRV as psychophysiological measurements of mental Workload. While driving, longitudinal and lateral measures were taken to assess driving performance. Every interaction with the display was recorded, as well as the latency of task completion as indicators of secondary task performance. At the end of the study, participants completed a demographic as well as a UX questionnaire (UEQ+; \cite{schrepp2019modulare}). 

\section{preliminary results}

Looking at the difference in workload in the city and the countryside environment in the no-task condition, we found that it was not significantly higher than zero for both HR and HRV. As the trend in the differences suggested effects in the different directions (\textit{M}$_{Diff}$(HR) = -1.28, \textit{M}$_{Diff}$(HRV) = 1.76), we performed exploratory one-sample two-sided \textit{t}-tests. The marginally non-significant result for HR (\textit{t}(32) = 1.93, \textit{p} = .063, \textit{d} = -0.34) suggests a decrease of MWL when going from the countryside to the city environment. This shows that the manipulation was ineffective, which could have an unexpected impact on the effects of our hypotheses. However, we still want to explore post-hoc if the adaption of the interface reduces the MWL in the city even further and consequently emphasizes the training effect. 

\begin{figure}[t]
	\centering
    \begin{subfigure}[t]{0.46\textwidth}
         \centering
         \includegraphics[width=6cm,height=4cm,keepaspectratio]{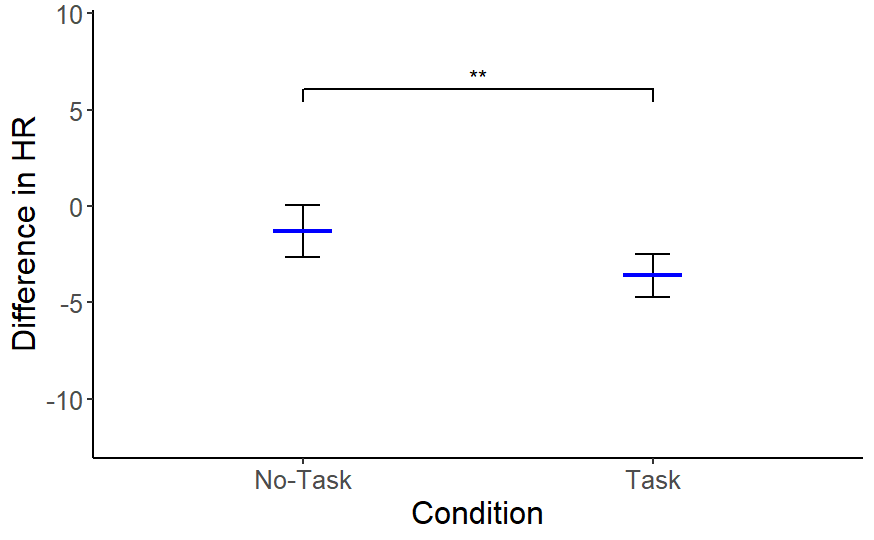}
         \caption{Heart Rate}
         \Description{Difference in HR}
         \label{fig:HR}
     \end{subfigure}
     \begin{subfigure}[t]{0.46\textwidth}
         \centering
         \includegraphics[width=6cm,height=4cm,keepaspectratio]{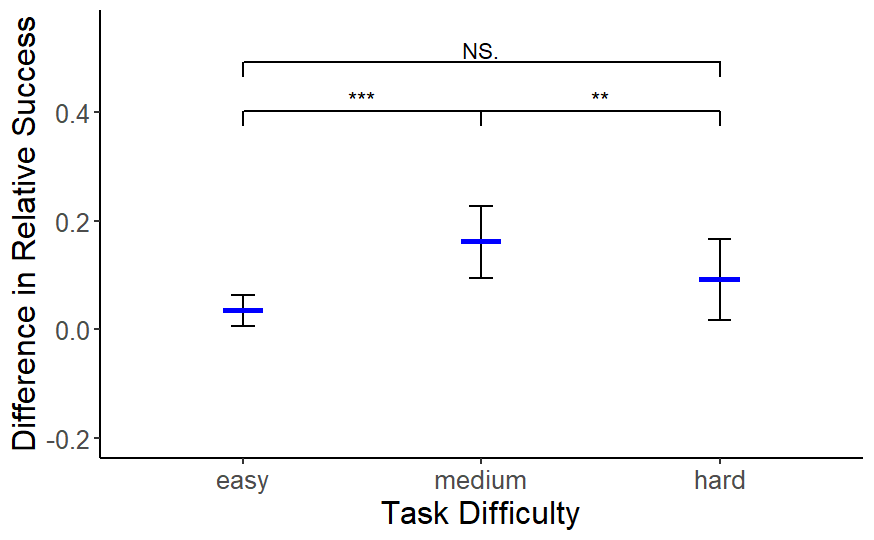}
         \caption{Relative Success}
         \Description{Difference in relative successes}
         \label{fig:Rel_suc}
     \end{subfigure}
     \begin{subfigure}[t]{0.46\textwidth}
         \centering
         \includegraphics[width=6cm,height=4cm,keepaspectratio]{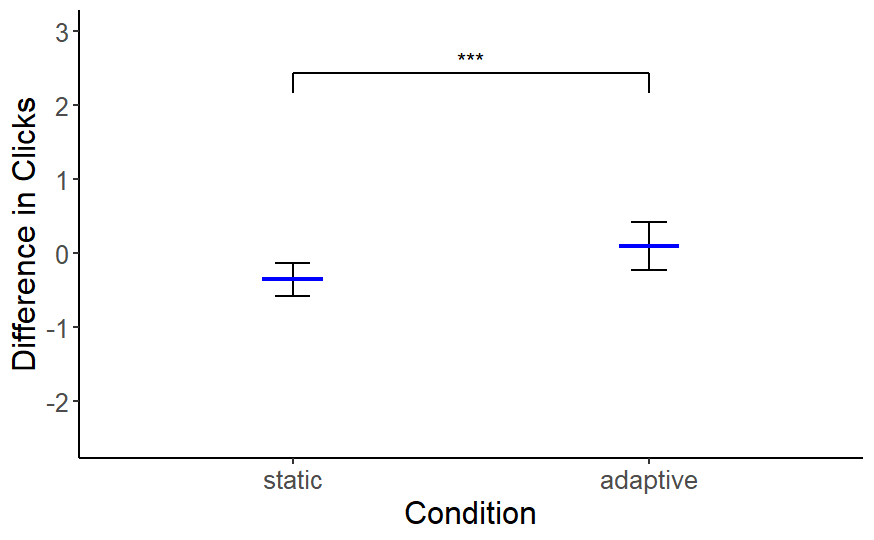}
         \caption{Number of clicks}
         \Description{Difference in number of clicks}
         \label{fig:Num_clicks}
     \end{subfigure}
     \caption{Preliminary results. For all plots, the y-axis represents the difference between the city and the countryside (city - countryside). (a) Participants showed a higher difference in HR in the no-task-condition than in the task conditions. (b) Especially for tasks of medium difficulty, participants showed a greater difference in how often they were able to solve the tasks. (c) Participants in the static condition showed better performance than the adaptive group.}
     \Description{ Participants showed a higher difference in HR in the no-task-condition than in the task conditions. \ref{fig:Rel_suc}: Especially for tasks of medium difficulty, participants showed a greater difference in how often they were able to solve the tasks.}
     \label{fig:results}
\end{figure}

Looking at hypothesis H1, there was a main effect of adaptation (\textit{F}(1,27) = 4.84, \textit{p} = .036, $\eta$$_{p}$$^{2}$ = .152) for the number of clicks. However, opposed to our hypothesis, participants in the adaptive group (\textit{M} = 0.09, \textit{SD} = 1.11) had a higher difference in clicks than participants in the static group (\textit{M} = -0.36, \textit{SD} = 0.70). Therefore,  participants in the static condition showed better performance when going from the countryside to the city environment (see fig. \ref{fig:Num_clicks}). 

   When comparing MWL in the no-task-condition and the task conditions we found a significant effect of task difficulty (\textit{Pillai-trace} = .106, \textit{F}(2,61) = 3.62, \textit{p} = .033, $\eta$$_{p}$$^{2}$ = .106). This effect was significant only for the difference in HR data (\textit{F} = 7.27, \textit{p} = .009). However, opposed to hypothesis H2, participants showed a lower difference in HR in the task conditions (\textit{M} = -3.59, \textit{SD} = 3.10) than in the no-task-condition (\textit{M} = -1.28, \textit{SD} = 3.82) (see fig. \ref{fig:HR}) 

Participants showed some performance differences for the interface tasks depending on the difficulties of the tasks, as hypothesis H3 predicted. We found significant main effects of task difficulty on the latency (\textit{F}(1.21, 32.76) = 15.71, \textit{p} > .001, $\eta$$_{p}$$^{2}$ = 0.368), on the number of successes (\textit{F}(1.69, 45.56) = 6.22, \textit{p} = .006, $\eta$$_{p}$$^{2}$ = 0.187) and on the relative number of successes (\textit{F}(1.58, 42.7) = 4.05, \textit{p} = .033, $\eta$$_{p}$$^{2}$ = .130). However, the direction of this effect shows no distinct trend (see fig. \ref{fig:Rel_suc}).

As opposed to hypothesis H4, the adaptive and the static group did not differ significantly in their UX ratings of the interface. The ratings were also non-significant when they were weighted with the importance ratings for the different items. 

\section{Discussion}
The analyses of our data showed a trend opposite to our predictions. Instead of showing decrements in performance and an increase in MWL we found a training effect. Possible explanations are a sequence effect leading to more experience with operating the driving simulator in the city environment, in addition to insufficient differences in environmental difficulty. Even though there have been attempts to establish rules~\cite{FASTENMEIER2007drivingtaskanalysis}, there is (as far as we know) no standard way of manipulating map design to ensure differences in driving difficulty~\cite{bobermin2021driving, papantoniou2017review}. Future research should focus on untangling those processes, which would make driving research more standardized and therefore lead to clearer results. Additionally, there should be a focus on long-term HMI interaction to avoid learning effects.

Another reason for our findings might be a speed reduction when performing tasks in the city. This is a common compensatory strategy adapted by drivers to help lower their MWL and to improve their performance in driving or other tasks~\cite{paxion2014mental}. This could explain the lower MWL in the city, and also why participants showed a lower difference in HR in the task condition than in the no-task condition. As speed reduction is a strategy more often adapted by experienced drivers~\cite{paxion2014mental}, it would also offer an alternative explanation to why there was a greater task performance improvement in the static condition, as indicated by the lower number of clicks. The static group may have been more experienced in driving, indicated by more kilometers driven per year. In general, it can be said that it is of high importance to further analyze the driving data before concluding, which is planned for our future work. 

Contrary to our predictions, UX ratings showed no clear preference for the static or the adaptive design. The display adaption even led to a smaller learning effect, as indicated by the higher difference of clicks in the adaptive group. This suggests that the adaption of in-vehicle interfaces with respect to environmental complexity can confuse the driver. However, our results need to be treated with care. Due to time restrictions, we were only able to collect a limited set of data and participants with high variance in driven kilometers per year, pointing to a difference in the driving experience. This could have led to the static group being able to allocate more resources to the secondary task and adapt more compensatory strategies to lower their MWL. We will be able to provide further explanations after analyzing driving performance data.

Future research should focus on different ways of adapting HMIs in cars, paying attention to mental workload, driving performance, and UX. Those experiments should be inspired by basic considerations of HMI design, driving safety concerns, and human psychology to reach the best possible option to make the daily driving task as comfortable, effective, and safe as possible. 

\begin{acks}
This work is partially funded by the German Ministry of Education and Research (BMBF) under project APX-HMI (Grant Number: 01IS17043) and project CAMELOT (Grant Number: 01IW20008).

\end{acks}

\balance
\bibliographystyle{ACM-Reference-Format}
\bibliography{references}

\appendix


\end{document}